\documentclass{article}
\usepackage{microtype}
\usepackage[T1]{fontenc}
\usepackage[utf8]{inputenc}
\usepackage{ismir}
\usepackage{amsmath,amsfonts,cite,url}
\usepackage{graphicx}
\usepackage{color}
\usepackage[table]{xcolor}
\usepackage{enumitem}
\usepackage[bookmarks=false]{hyperref}
\hypersetup{
  colorlinks=true,
  linkcolor=black,
  citecolor=black,
  urlcolor=black,
}
\usepackage{booktabs}

\title{MAJEPPA: Morphing and Assessing in a Unified Piano Performance Space}

\multauthor
  {Jinwen Zhou$^1$ \hspace{1cm} Huan Zhang$^1$ \hspace{1cm} Weixi Zhai$^2$}
  {{\bf Jinhua Liang$^1$ \hspace{1cm} Aidan O. T. Hogg$^1$ \hspace{1cm} Simon Dixon$^1$}\\
  $^1$ Queen Mary University of London, UK\\
  $^2$ Quanzhou Normal University, China\\
  {\tt\small [jinwen.zhou, huan.zhang, jinhua.liang, a.hogg, s.e.dixon]@qmul.ac.uk$^1$, wishzhai@gmail.com$^2$}
  }

\def\authorname{J. Zhou, H. Zhang, W. Zhai, J. Liang, A. O. T. Hogg, and S. Dixon}

\begin{document}

\maketitle

\begin{abstract}
We present MAJEPPA, a self-supervised framework to learn piano performance representations that span the full skill spectrum, from beginner practice sessions to virtuoso concert recordings. We curate the MAJEPPA dataset, comprising ${\sim}$4,000 annotated recordings across six expertise levels and six recording contexts. We adapt a single pre-trained MIDI autoregressive model with a joint objective: next-token prediction learns score-conditioned performance generation at various skill levels, while InfoNCE and supervised contrastive losses align abstract score and performance representations in a joint embedding space. The proposed model both generates and understands performances in a unified framework. By introducing the EVPMR benchmark, a suite of downstream tasks spanning quality assessment, competition ranking, mistake and technique classification, we evaluate the learnt representations, demonstrating progress towards a real-world model for the piano performance space. 
\end{abstract}


\section{Introduction}\label{sec:introduction}

Piano performance is a remarkably rich domain of human expression. The same score can be realised as a child's halting first attempt, an adult student sight-reading, a teacher demonstrating phrasing at half tempo, or a virtuoso commanding the concert stage. Understanding this variation computationally, distinguishing not just \textit{what} is played but \textit{how}, is central to music education, where intelligent systems must diagnose mistakes~\cite{morsi2023sounds,morsi2024simulating, chang2025RUMAA}, assess quality~\cite{Zhang2024HowDataset,Zhang2024FromJudges}, provide formative feedback~\cite{zhang2025llaqo}, estimate difficulty~\cite{libricky2025saxophone,mullerschon2025playability}, and personalise learning~\cite{hasseinbey2025guitar} across student abilities. Recent efforts have begun capturing the learning process itself through rehearsal datasets~\cite{morsi2025rach3} and multimodal practice recordings~\cite{kim2025pianovam,choi2025flutist}. Yet existing approaches address these goals in isolation: task-specific models are trained on narrow datasets for individual objectives, yielding representations that do not transfer across tasks or skill levels.

We argue that a \textit{unified} representation of piano performance is more foundational, encoding core performance parameters from which higher-level structure is derived, and more useful in practice. However, building such a representation is hindered by two gaps in MIR research. First, a \textbf{data gap}: existing symbolic music datasets focus almost exclusively on professional performances~\cite{zhang_atepp_2022,foscarin2020asap}, leaving most real-world piano playing (practice, sight-reading, demonstrations) unrepresented. 
Second, a \textbf{methodological gap}: we need a learning framework that captures \textit{abstract} performance characteristics (style, accuracy, expressivity) alongside and linked to the score.
We address this gap with the Joint-Embedding Predictive Architecture (JEPA)~\cite{lecun2022path}, a framework that learns representations by predicting in latent space, focusing on high-level semantics. Furthermore, LLM-JEPA~\cite{huang2025llmjepa} demonstrates that predictive latent modelling with language models can yield representations that are not only abstract but also perceptually grounded.

Given a score, we want to predict the \textit{abstract character} of how that score will be realised. The score serves as the context view (specifying \textit{what} to play) and the performance as the target view (capturing \textit{how} it is played). The model must predict the abstract performance embedding from the score, along with descriptive conditioning (performer type, recording context), without observing the performance tokens. A joint generative objective prevents representation collapse and retains the model's ability to generate, resulting in a single model that both produces performances and yields transferable embeddings that specialise in understanding real-world performances.

To close the data gap, we curate the \textbf{MAJEPPA dataset}, comprising ${\sim}$4,000 annotated piano recordings spanning six expertise levels and six recording contexts. We evaluate our JEPA-based model on the Evaluation of Piano MIDI Representations (EVPMR) benchmark using frozen representations with linear probes. Our contributions are:
\begin{enumerate}[noitemsep]
    \item The \textbf{MAJEPPA dataset}, the first large-scale dataset to span the full performance spectrum from child beginners to virtuosi;
    \item The first \textbf{JEPA} to model for joint symbolic piano-performance generation and representation learning, using score–performance pairs as a natural two-view structure;
    \item The \textbf{EVPMR} benchmark suite, designed for evaluation across downstream tasks, including Chopin competition ranking, technique detection, performance quality regression, and conspicuous mistake prediction, where our models outperform various baselines.
\end{enumerate}

\section{Related Work}

\subsection{Music and MIDI representation learning}

Approaches to learning representations from symbolic music are shaped by how MIDI is serialised into a sequence, and tokenisation schemes such as REMI~\cite{huang2020remi}, Compound Word~\cite{hsiao2021cpword}, OctupleMIDI~\cite{zeng2021musicbert}, and PerTok~\cite{lenz2024pertok} provide the input vocabularies on top of which the following families of self-supervised objectives are applied.

Masked-reconstruction pre-training: MusicBERT~\cite{zeng2021musicbert} and MidiBERT-Piano~\cite{midibertpiano} apply BERT-style masked token prediction to symbolic music, producing representations that transfer to composition-level classification (genre, emotion, melody extraction). PianoBART~\cite{min2025pianobart} extends this with BART-style denoising and multi-level masking over Octuple-encoded piano MIDI, unifying understanding and generation in a single encoder-decoder.

Autoregressive foundation models: A parallel line of work treats MIDI as a language and trains large causal transformers. ChatMusician~\cite{yuan2024chatmusicianunderstandinggeneratingmusic} and MuseCoco~\cite{lu2023musecocogeneratingsymbolicmusic} target controllable composition from text, while Moonbeam~\cite{guo2025moonbeammidifoundationmodel} scales next-token prediction to 18\,B MIDI tokens, pre-training on ${\sim}$81,600 hours of data using a Fundamental Music Embedding tokenizer and multidimensional relative attention. Aria~\cite{bradshaw2025aria} focuses on solo piano, pre-training on ${\sim}$60,000 hours of MIDI and obtaining strong embeddings via subsequent contrastive fine-tuning.

Contrastive and cross-modal alignment: CLaMP\,3~\cite{wu2025clamp3universalmusic} aligns multilingual text, symbolic music (ABC and MIDI), and audio in a shared embedding space through contrastive learning, supporting retrieval across modality pairs that were never jointly observed during training. In the piano-performance setting, Pianist Transformer~\cite{you2025pianist} removes the need for score-performance alignment and pre-trains on large unaligned corpora for expressive rendering.

Downstream evaluation of these representations has centred on composition-level classification (genre, emotion, chord recognition). Performance-level understanding, in contrast, remains fragmented across isolated studies, including difficulty estimation~\cite{ramoneda2023cipi}, pianist identification~\cite{tang2023pianistid}, and score-informed mistake detection~\cite{chou2026laddersym}. Building on these downstream evaluation tasks and inspired by the Evaluation Package for Audio Representations (\href{https://github.com/nttcslab/eval-audio-repr}{\textit{EVAR}}), the unified EVPMR benchmark is introduced in this work.

\subsection{Joint-Embedding Predictive Architecture (JEPA)}
JEPA~\cite{lecun2022path} learns representations by predicting target latent embeddings from partial context without reconstruction or contrastive objectives. Using an asymmetric encoder–predictor design, it focuses on modelling predictable structure in latent space, enabling abstract and semantically meaningful representations across domains.
This framework has been used to learn semantic image representations by predicting masked spatial target blocks from context (I-JEPA~\cite{assran2023self}), and spatio-temporal representations by predicting latent targets from video context (V-JEPA 2~\cite{assran2025vjepa2selfsupervisedvideo}). Audio-JEPA~\cite{tuncay2025audio} translates the JEPA principle to the audio domain by predicting latent representations of masked mel-spectrogram patches.

In the music domain, Stem-JEPA~\cite{StemJEPA} leverages the multi-track structure of music to replace input-space masking with stem holdout, training the predictor to predict the latent representation of a missing instrument from a context mix conditioned on instrument labels. JEP-AGA~\cite{hu2025compose} combines the principles of JEPA with autoregressive generation for music infilling. AMT-JEPA~\cite{pilataki2025jepa_amt}
demonstrates that JEPA can learn semantically meaningful representations for multi-instrument music transcription, particularly capturing instrument and pitch-related information. Hachana and Rasheed~\cite{hachana2025using} extend JEPA to symbolic domain with music-specific masking strategies for MIDI sequences.

\begin{figure}[t]
    \centering
    \includegraphics[trim=0.5cm 9cm 0.5cm 0.5cm, width=\linewidth]{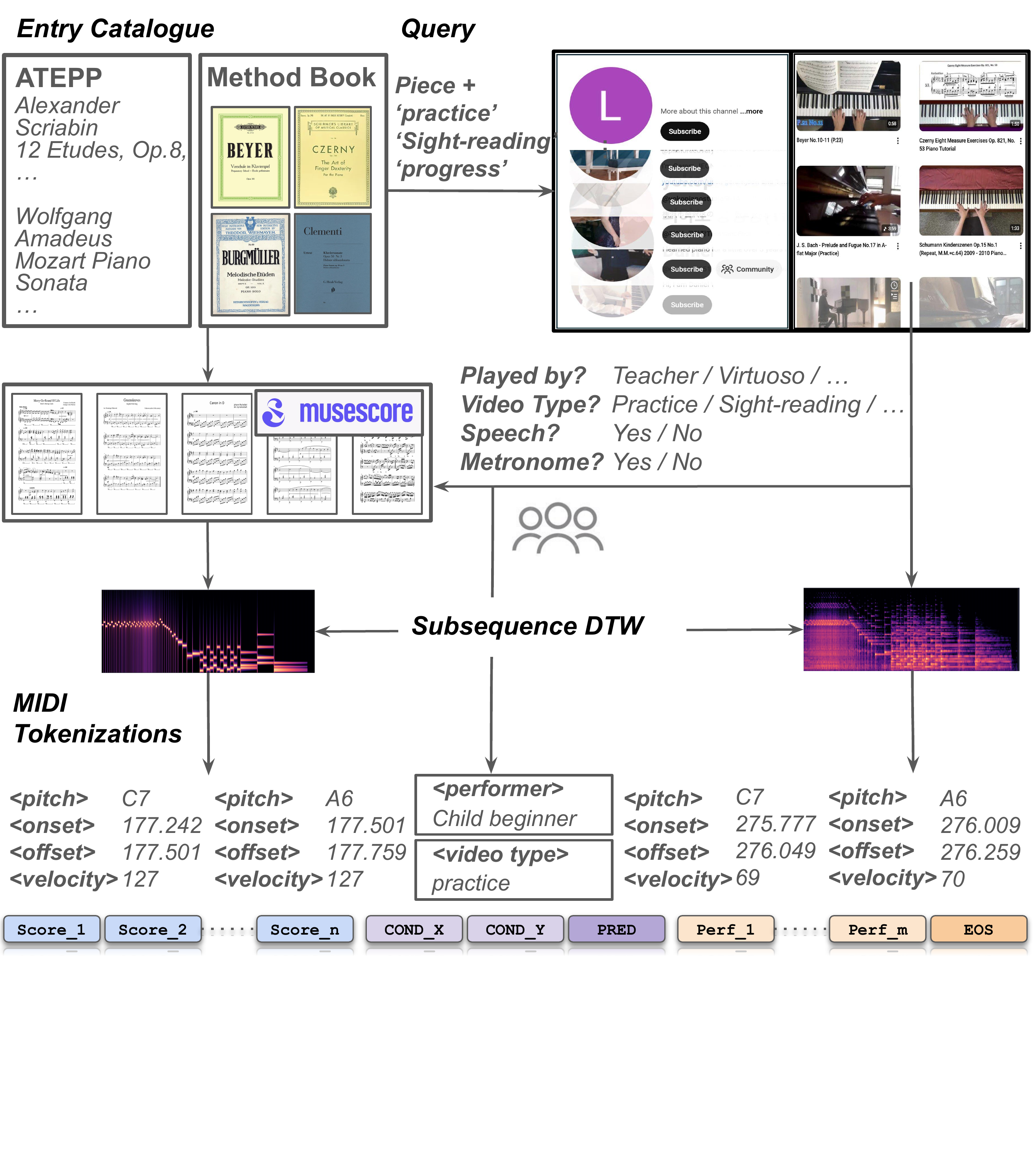}
    \caption{Data curation pipeline. YouTube recordings are collected from entry catalogues and search queries, annotated with performer expertise and recording context, aligned with the corresponding score via subsequence DTW. Aligned pairs are segmented and tokenised for pre-training.}
    \label{fig:data_proc}
\end{figure}

\section{Data Curation}\label{sec:data}

\begin{figure}
    \centering
    \includegraphics[trim=0.5cm 0.5cm 0cm 0cm, width=\linewidth]{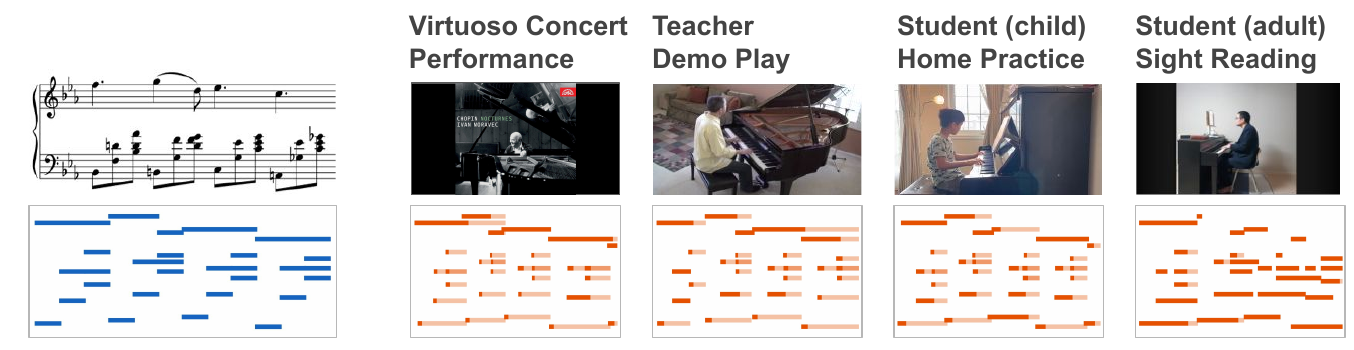}
    \caption{The same piece realised across different expertise levels and recording contexts. Left: score excerpt with MIDI piano roll. Right: four performances showing visible differences in timing, note accuracy, and dynamics.}
    \label{fig:data_demo}
\end{figure}

Existing symbolic music datasets focus predominantly on professional performances: ATEPP~\cite{zhang_atepp_2022} contains virtuoso interpretations and ASAP~\cite{foscarin2020asap} provides competition-level recordings with aligned scores. To address this gap, we curate the \textbf{MAJEPPA dataset}, comprising 3,979 real-world piano recordings across 942 distinct pieces and movements, spanning the full performance spectrum from child beginners to concert virtuosi (\figref{fig:data_demo}).

\textbf{Collection}: Solo-piano recordings that are publicly available are collected from YouTube, covering demo recordings, informal practice, and sight-reading sessions. To ensure diversity of performance levels across the music pieces, we source our data from catalogues and channels (top of \figref{fig:data_proc}).
Firstly, we query for repertoire titles, adding keywords such as \textit{Practice, Sight-reading}, or \textit{Progress}.
For concert-repertoire level pieces, we query for videos of the same pieces or movements as those in ATEPP.
To include beginner-level material, we query for exercises and pedagogical repertoire commonly used in piano training, such as method books by Carl Czerny, Friedrich Burgm\"{u}ller, and Anton Diabelli. We also include widely practised contemporary pieces such as \textit{Kiss the Rain} by Yiruma. 
From the results, we manually identify resourceful channels from piano teachers (demonstrations and tutorials), sight-reading sessions, practice recordings, and learning progress videos.

\subsection{Annotation}\label{subsec:annotation}

After sourcing videos, each recording is manually annotated by conservatoire student pianists\footnote{\label{fn:supp}Annotator reliability: \url{https://jepa-demo.vercel.app}} who work from a shared calibration document that defines each category and provides reference clips. 
The annotations capture differences along two orthogonal axes: recording context and performer expertise (Table \ref{tab:distribution}). Recording context describes the setting in which the performance was captured:

\begin{itemize}[noitemsep]
    \item \textbf{Practice}: Recordings from students who are actively learning a particular piece. 
    \item \textbf{Sight-reading}: Sight-reading sessions involving one or multiple unfamiliar pieces. 
    \item \textbf{Showcase}: Recordings presented to teachers (e.g., prior to weekly lessons) or in student concerts, reflecting a more polished, near-standard performance with relatively fewer mistakes despite the performer’s developing skill level. 
    \item \textbf{Demo}:  Demonstration recordings by piano teachers illustrating how a piece should be performed. They are usually highly accurate in terms of pitch and timing, and not overly expressive.  
    \item \textbf{Slow demo}: Slow-tempo demonstration recordings by piano teachers.
    \item \textbf{Concert}: Commercial recordings or live performances from pianists.
\end{itemize}

Performer expertise describes the background of the performer:
\begin{itemize}[noitemsep]
    \item \textbf{Child beginner}: Young learners performing simple exercises or early-stage repertoire. 
    \item \textbf{Child professional}: Young performers demonstrating advanced technical proficiency, typically on a pre-professional or professional training track. 
    \item \textbf{Adult beginner}: Adult learners performing introductory exercises or beginner repertoire, typically engaging with piano as a hobby. 
    \item \textbf{Adult intermediate}: Performers demonstrating moderate technical proficiency and developing musicality. 
    \item \textbf{Piano teacher}: Instructors providing demonstration performances for instructional purposes. 
    \item \textbf{Virtuoso}: Established professional pianists.
\end{itemize}

Besides the conditioning tags, the annotators also provide the additional information: 
\textbf{1). Breakpoints} for long videos, especially when multiple pieces are in a single recording;
\textbf{2). Corresponding scores} sourced from \href{https://musescore.com}{\textit{MuseScore}}
for each identified piece, with annotators visually matching the score edition to the performed arrangement (e.g., key, structure, and approximate length);
\textbf{3). Filtering out} videos that contain speech, metronome, or other non-piano sounds, and those not meeting the dataset criteria. 


\begin{table}[t]
\caption{Distribution of recordings by performer expertise (rows) and recording context (columns).}
\label{tab:distribution}
\centering
\setlength{\tabcolsep}{3pt}
\footnotesize
\begin{tabular}{lrrrrrr|r|r}
\toprule
 & Prac. & Sight. & Show. & Demo & Slow & Conc. & \textbf{Total} & \textbf{Hours}\\
\midrule
Teacher    & 35  & 15  & 55  & 690 & 311 & 173 & 1279 & 76.1\\
Ad.\ Beg.  & 622 & 217 & 92  & 12  & 3   & 1   & 947  & 127.9\\
Ad.\ Int.  & 444 & 135 & 177 & 16  & 2   & 3   & 777  & 50.7\\
Ch.\ Beg.  & 186 & 8   & 193 & 2   & 2   & 0   & 391  & 8.9\\
Ch.\ Pro.  & 84  & 0   & 161 & 2   & 0   & 0   & 247  & 13.7\\
Virtuoso   & 0   & 0   & 0   & 0   & 0   & 338 & 338  & 28.9\\
\midrule
\textbf{Total} & 1371 & 375 & 678 & 722 & 318 & 515 & \textbf{3979} & 303.2\\
\bottomrule
\end{tabular}
\end{table}

\begin{figure*}[t]
  \centering
  \includegraphics[
  alt={Sequence packing and attention masks. View~1 (score, conditioning, and predictor tokens) and View~2 (performance tokens) are packed into a single sequence. Item~1 uses a standard causal mask; Item~2 uses a block-diagonal mask where the two views cannot attend to each other.},
  trim=0cm 1.2cm 0cm 0cm, width=0.9\linewidth]{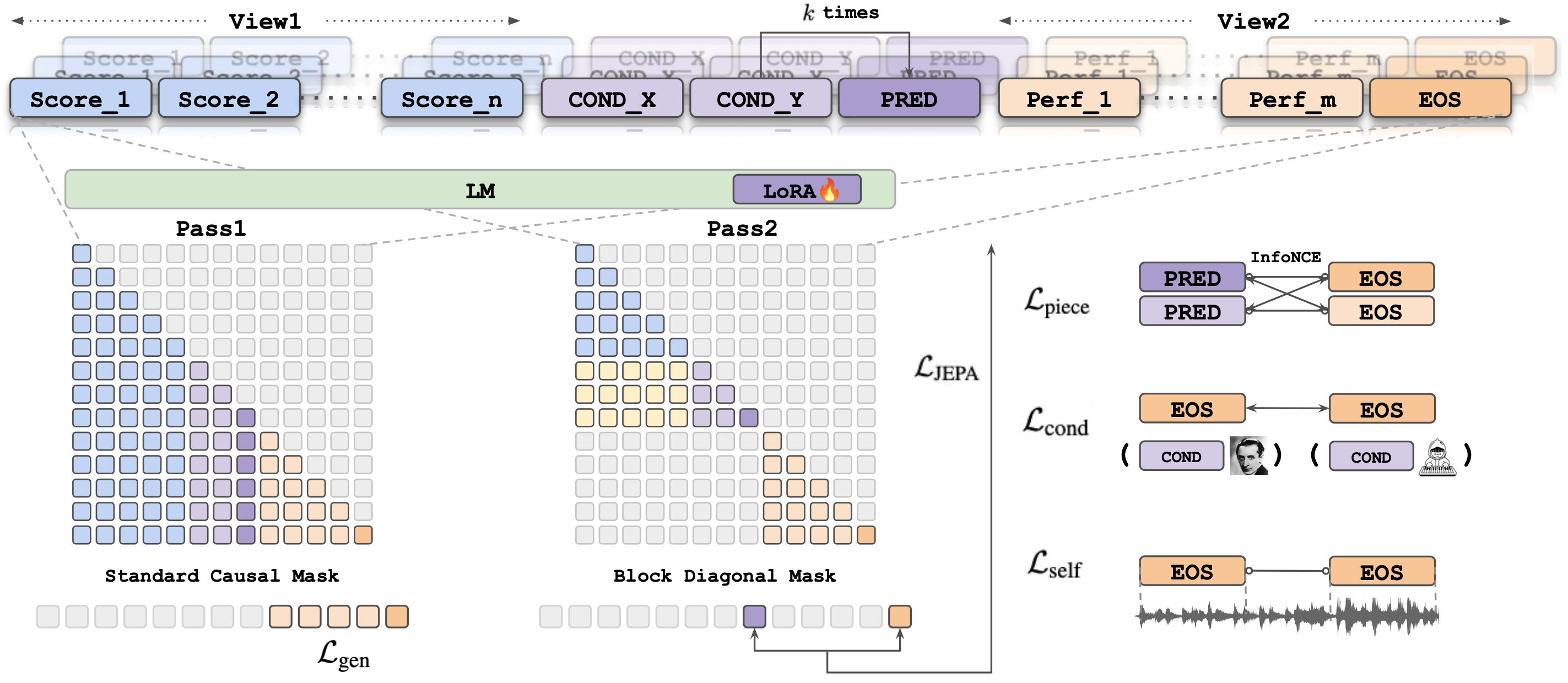}
  \caption{Sequence packing and attention mask design. \textbf{Left:} standard causal mask for the generative item. \textbf{Middle:} block-diagonal causal mask for the JEPA item, enforcing independent encoding of score and performance views. \textbf{Right:} contrastive objectives for piece alignment, cross-condition clustering, and within-performance alignment.}
  \label{fig:mask}
\end{figure*}

\subsection{Processing}\label{subsec:processing}

The final processing pipeline produces aligned pairs of score and performance MIDI segments: 
\textbf{1). Breakpoint processing}: long recordings are first segmented into individual performance excerpts based on annotation.
Each segment is associated with its corresponding score; 
\textbf{2). Transcription}: performance audio is transcribed into MIDI using Transkun~\cite{yan_scoring_2024}, chosen as the note-level state of the art on standard piano AMT benchmarks rather than alternatives such as Aria-AMT~\cite{bradshaw2024ariaamt}, though we did not explicitly check transcription quality (including pedal artefacts);
\textbf{3). Alignment}: to enable segment-level pairing under the limited context length of language models, score and performance must be temporally aligned before cropping. Due to the highly non-ideal and fuzzy nature of real-world data (e.g., mistakes, omitted passages, one-hand practice), we perform subsequence dynamic time warping (DTW) on chroma features extracted from synthesised score audio and performance audio, allowing robust partial alignment.
Subsequence DTW handles omitted repeats by aligning the performance to the relevant contiguous score portion; performances with extra repetitions may produce locally noisier pairs, which the segment-level objective tolerates. The resulting warping path maps score time to performance time, enabling consistent segmentation into fixed-token-length windows for downstream processing.

\section{Proposed Method}\label{sec:jepa_pretrain}
Our goal is to learn a unified representation of piano performance that captures both the underlying musical content and its expressive realisation. To this end, we model score and performance as two complementary views of the same musical excerpt, and learn to predict performance representations from score representations in a shared latent space.

We formulate each training example as a \textit{score--performance pair} $(\mathbf{s}, \mathbf{p})$ corresponding to the same musical excerpt. The score view $\mathbf{s}$ is a deadpan tokenisation of the written score, while the performance view $\mathbf{p}$ is the tokenised expressive realisation of the same music. Both share the same musical content, but differ in the performance detail: the score specifies the intended notes and structure, whereas the performance carries timing, dynamics, articulation, and recording context. Following LLM-JEPA~\cite{huang2025llmjepa}, we treat this pair as a natural two-view structure and train a predictor to map from the score view to the performance view in representation space. A mini-batch contains $B$ such pairs.

The objective is to learn a unified representation of piano performance that supports both conditional generation and abstract embedding prediction. Specifically, the model is trained to (i) generate performance tokens conditioned on score tokens, and (ii) predict a latent representation of the performance from the score and conditioning tokens alone. 

\subsection{Model components}
We implement this two-view predictive framework within a single autoregressive transformer by augmenting it with specialised tokens and projection heads. 
The architecture comprises an encoder backbone, a predictor via \texttt{[PRED]} tokens, conditioning tokens encoding performer and recording context, and a projection head for contrastive objectives.

\textbf{Encoder.} We use Aria-medium~\cite{bradshaw2025aria}, a 632M-parameter LLaMA-style autoregressive transformer pre-trained on symbolic MIDI. \textbf{Predictor.} We append $k$ \texttt{[PRED]} tokens to the input sequence and reuse the model’s self-attention as a tied-weights predictor: the final \texttt{[PRED]} state represents $\mathrm{Pred}(\mathbf{s})$. \textbf{Condition tokens.} Two tokens, \texttt{[COND\_X]} (performer) and \texttt{[COND\_Y]} (recording context), provide 
categorical conditioning via learned embeddings (Section~\ref{sec:data}).

\subsection{Sequence construction and attention masking}
Each score--performance pair $(\mathbf{s}, \mathbf{p})$ is first packed into a single token sequence as shown in Fig.~\ref{fig:mask}:
$
[\mathbf{s};\ \texttt{COND};\ \texttt{PRED};\ \mathbf{p};\ \texttt{EOS}],
$
where $\mathbf{s}$ denotes the score tokens, $\mathbf{p}$ the performance tokens, and \texttt{COND}, \texttt{PRED}, and \texttt{EOS} are special tokens marking the conditioning boundary, prediction query, and sequence termination, respectively.


\textbf{Pass~1 -- generative (causal mask).} The packed sequence $[\mathbf{s};\, \texttt{COND};\, \texttt{PRED};\, \mathbf{p};\, \texttt{EOS}]$ is processed under a standard lower-triangular causal mask, so performance tokens attend to the full preceding context. This pass produces the next-token logits over the performance tokens used by the generative loss $\mathcal{L}_{\text{gen}}$ defined below.

\textbf{Pass~2 -- JEPA (block-diagonal causal mask).} The same tokens are processed under a block-diagonal mask that defines two independent views: View~1 $[\mathbf{s};\, \texttt{COND};\, \texttt{PRED}]$ and View~2 $[\mathbf{p};\, \texttt{EOS}]$, each internally causal but mutually invisible. We extract $\mathrm{Pred}(\mathbf{s})$ from the \texttt{[PRED]} position in View~1 and $\mathrm{Enc}(\mathbf{p})$ from the \texttt{[EOS]} position in View~2. By construction (Fig.~\ref{fig:mask}), \texttt{[PRED]} attends only to score and conditioning tokens.

This split ensures (i)~Pass~1 retains the model's ability to generate performances conditioned on scores, and (ii)~Pass~2 must form an abstract prediction from the score alone, since \texttt{[PRED]} never observes the performance tokens.

\subsection{Training objective}\label{subsec:objective}
The training loss combines a generative term with contrastive terms that align score and performance representations, enforce condition-aware structure, and maintain within-performance consistency. For each pair $i$ in the batch, we obtain two embeddings via a shared linear projection head $\phi: \mathbb{R}^{1536} \to \mathbb{R}^{512}$, both $\ell_2$-normalised:
\begin{equation*}
\mathbf{z}_i^s = \phi(\mathrm{Pred}(\mathbf{s}_i)) \quad\text{and}\quad \mathbf{z}_i^p = \phi(\mathrm{Enc}(\mathbf{p}_i)),
\end{equation*}
where $\mathrm{Pred}(\mathbf{s}_i)$ and $\mathrm{Enc}(\mathbf{p}_i)$ are the Pass~2 outputs from the \texttt{[PRED]} and \texttt{[EOS]} positions respectively. The training loss combines a generative term computed from Pass~1 with three contrastive JEPA terms computed from Pass~2:
\begin{equation}\label{eq:loss}
\mathcal{L} \;=\; \gamma\,\mathcal{L}_{\text{gen}} \;+\; \lambda\,\big(\mathcal{L}_{\text{piece}} + \beta\,\mathcal{L}_{\text{cond}} + \alpha\,\mathcal{L}_{\text{self}}\big),
\end{equation}
where the four loss terms are defined below and $\gamma, \lambda, \beta, \alpha \geq 0$ are scalar weights. All contrastive terms share a single temperature $\tau$.

\textbf{Generative loss} $\mathcal{L}_{\text{gen}}$ is the standard next-token cross-entropy over the performance tokens of $\mathbf{p}$, conditioned on score and conditioning tokens through the Pass~1 causal mask.

\textbf{Piece alignment} $\mathcal{L}_{\text{piece}}$ is an InfoNCE loss that pulls each predicted score embedding towards its paired performance embedding within the batch:
\begin{equation}\label{eq:infonce}
\mathcal{L}_{\text{piece}} = -\frac{1}{B}\sum_{i=1}^{B} \log \frac{\exp(\mathbf{z}_i^s \cdot \mathbf{z}_i^p / \tau)}{\sum_{j=1}^{B} \exp(\mathbf{z}_i^s \cdot \mathbf{z}_j^p / \tau)}.
\end{equation}

\textbf{Cross-condition clustering} $\mathcal{L}_{\text{cond}}$ is a soft supervised contrastive loss over performance embeddings. Let $w_{ij} \in [0,1]$ denote a continuous similarity between samples $i$ and $j$, derived from the ordinal distance between their performer skill levels and recording formality levels, and let $\mathcal{C}_i = \{j : \text{cond}(j) \neq \text{cond}(i)\}$ restrict to pairs with different condition tags:
\begin{equation}\label{eq:supcond}
\mathcal{L}_{\text{cond}} = -\frac{1}{B}\sum_{i=1}^{B} \sum_{j \in \mathcal{C}_i} \bar{w}_{ij} \log \frac{\exp(\mathbf{z}_i^p \cdot \mathbf{z}_j^p / \tau)}{\sum_{h \in \mathcal{C}_i} \exp(\mathbf{z}_i^p \cdot \mathbf{z}_h^p / \tau)},
\end{equation}
with $\bar{w}_{ij} = w_{ij} / \sum_{h \in \mathcal{C}_i} w_{ih}$ row-normalised over $\mathcal{C}_i$. This term replaces the cosine-distance loss of~\cite{huang2025llmjepa}, which we found prone to representation collapse in our setting. See our website\footnotemark[1] for details of continuous condition similarity design.

\textbf{Within-performance alignment} $\mathcal{L}_{\text{self}}$ is a symmetric InfoNCE loss over pairs of segments from the same performance. We form $B$ pairs $(\mathbf{p}_i^{a}, \mathbf{p}_i^{b})$ with embeddings $\mathbf{z}_i^{a}, \mathbf{z}_i^{b}$, and define
\begin{align}\label{eq:self}
\mathcal{L}_{\text{self}} = -\frac{1}{B} \sum_{i=1}^{B} \sum_{u \in \{a,b\}} 
\log \frac{\exp(\mathbf{z}_i^{u}\!\cdot\!\mathbf{z}_i^{\bar{u}}/\tau)}
{\sum_{j=1}^{B} \exp(\mathbf{z}_i^{u}\!\cdot\!\mathbf{z}_j^{\bar{u}}/\tau)},
\end{align}
where $\bar{a}=b$ and $\bar{b}=a$. 
Both terms encourage segments from the same performance to be more similar than segments from different recordings, while averaging over the two directions yields a symmetric objective. Since the two segments share performer and recording tags, $\mathcal{L}_{\text{self}}$ complements $\mathcal{L}_{\text{cond}}$ by encouraging within-performance coherence.


\textbf{Training setup: }
We freeze the pre-trained Aria weights and apply Low-Rank Adaptation (LoRA)~\cite{hu2022lora} to the attention projections \texttt{mixed\_qkv} and \texttt{att\_proj\_linear} (rank $r\!=\!512$, scaling $\alpha_{\text{LoRA}}\!=\!32$), giving $\sim$75M trainable parameters out of 660M. The context window is extended from 2048 to 4096 tokens via linear RoPE scaling~\cite{su2024roformer}. Loss hyperparameters in Eq.~\eqref{eq:loss} are set to $\gamma\!=\!1$, $\lambda\!=\!8$, $\beta\!=\!1$, $\alpha\!=\!1$, with temperature $\tau\!=\!0.07$ and $k\!=\!1$ \texttt{[PRED]} token, chosen by ablation. 

We train on 4$\times$NVIDIA A100-80GB GPUs in bfloat16 mixed precision for 120 epochs, with a per-GPU batch size of 8 and gradient accumulation of 4. Optimisation uses AdamW (weight decay 0.01) with separate learning rates for these LoRA modules ($2{\times}10^{-4}$), the new conditioning and \texttt{[PRED]} token embeddings ($1{\times}10^{-3}$), and the LM head ($1{\times}10^{-5}$).

\begin{table*}[t]
  \centering
  \caption{Downstream evaluation on the EVPMR benchmark. 
  Mean (std.) over 5-fold cross-validation across 5 random seeds.
  }
  \label{tab:evpmr}
  \footnotesize
  \setlength{\tabcolsep}{4pt}
  \begin{tabular}{l c c ccc ccc c}
  \toprule
  & {Chopin}
  & {Technique}
  & \multicolumn{3}{c}{Quality Regression: MAE$\downarrow$}
  & \multicolumn{3}{c}{UMP: $F1$}
  & {Pianist8} \\
  \cmidrule(lr){2-2}\cmidrule(lr){3-3}\cmidrule(lr){4-6}\cmidrule(lr){7-9}\cmidrule(lr){10-10}
  Model & PairAcc & Acc
        & PISA & NeuroPiano & YCU-PPE
        & EN & BM & SI
        & Acc \\
  \midrule
  Aria
    & 47.5\,(8.1) & 66.2\,(11.4)
    & .122\,(.031) & .215\,(.019) & .068\,(.005)
    & .174\,(.025) & .313\,(.016) & .300\,(.020)
    & 86.1\,(4.3) \\
  Moonbeam
    & 53.7\,(2.9) & 77.9\,(4.8)
    & .151\,(.035) & .192\,(.014) & .061\,(.004)
    & -- & -- & --
    & 80.6\,(4.5) \\
  MusicBERT
    & 52.5\,(5.8) & 67.1\,(4.6)
    & .190\,(.026) & .205\,(.019) & .065\,(.004)
    & -- & -- & --
    & 78.6\,(2.1) \\
  CLaMP\,3
    & 53.5\,(8.5) & 70.3\,(2.7)
    & .217\,(.036) & .203\,(.017) & .089\,(.004)
    & -- & -- & --
    & 80.8\,(3.3) \\
  \midrule
  MAJEPPA ($\mathcal{L}_\text{piece}$)
    & 53.7\,(8.1) & 76.3\,(5.9)
    & .125\,(.083) & .180\,(.015) & .064\,(.004)
    & .151\,(.021) & .308\,(.012) & .313\,(.015)
    & 83.1\,(3.2) \\
  MAJEPPA ($\mathcal{L}_\text{piece}$+$\mathcal{L}_\text{cond}$)
    & 58.7\,(8.7) & 70.3\,(9.5)
    & .108\,(.090) & .181\,(.022) & .062\,(.005)
    & .141\,(.015) & .321\,(.012) & \textbf{.341}\,(.004)
    & \textbf{86.8}\,(5.2) \\
  MAJEPPA ($\mathcal{L}_\text{piece}$+$\mathcal{L}_\text{cond}$+$\mathcal{L}_\text{self}$)
    & \textbf{59.7}\,(9.4) & \textbf{79.0}\,(6.5)
    & \textbf{.090}\,(.034) & \textbf{.155}\,(.018) & \textbf{.059}\,(.002)
    & \textbf{.179}\,(.012) & \textbf{.333}\,(.015) & .316\,(.037)
    & 84.2\,(3.8) \\
  \bottomrule
  \end{tabular}
  \end{table*}

\section{Downstream tasks and evaluation}
Inspired by EVAR, we formulate performance-related tasks in a unified package, EVPMR,
providing standardised evaluation protocols. All audio-only datasets are transcribed to MIDI with Transkun~\cite{yan_scoring_2024} for consistency. Every task is evaluated with 5-fold cross-validation repeated across 5 random seeds. On top of a frozen global embedding from \texttt{encode()}, classification uses a linear probe, pairwise ranking uses a binary classification over concatenated pair features, regression uses Ridge, and UMP uses a frame-level MLP decoder.

\textbf{Performance assessment.} Three datasets provide normalised ratings in $[0, 1]$: NeuroPiano~\cite{Zhang2024HowDataset} (104 recordings, 27 students, 6 pieces, rated by 33 teachers on a 0--6 scale), YCU-PPE-III~\cite{Wang2021Audio-basedMechanism} (606 performances, 13 songs, rated by 3 instructors on a 0--100 scale), and PISA~\cite{parmar2021pianoskillsassessment} (59 YouTube videos, skill level 1--10). Ratings are aggregated per recording and normalised using min–max scaling. We report MAE$\downarrow$ as the primary metric.

\textbf{Chopin competition ranking.} Following PianoJudge~\cite{Zhang2024FromJudges}, this task uses 130 preliminary-round performances from the 2015 International Chopin Piano Competition, each labelled with the round the performer reached (0--4). We cast it as pairwise ranking: for every ordered pair of performances with different labels, the probe predicts which of the two advanced further. We report pairwise ranking accuracy (chance = 0.50).

\textbf{Technique detection.} From the Pianism-Labelling Dataset~\cite{Zhang2024FromJudges} we use the single-label piano technique classification task, which assigns each performance excerpt to one of 7 technique categories (octaves, scales, etc.).

\textbf{Pianist8.} Introduced with MidiBERT-Piano~\cite{midibertpiano}, this dataset contains 411 pieces by 8 pianists spanning four stylistic categories.
The task is 8-way pianist identification, testing whether representations capture individual styles.

\textbf{Unreferenced Mistake Prediction (UMP).} Following the score-free ``conspicuous mistake'' formulation introduced by~\cite{morsi2023sounds}, UMP detects performance errors from musical context alone. We evaluate on three open-source datasets: EN-augmented $(\textit{EN})$~\cite{Jiang2023ExpertFeedback}, Burgm\"uller $(\textit{BM})$~\cite{morsi2023sounds}, and ScoreInformed $(\textit{SI})$~\cite{benetos2012score}. UMP is token-level: we extract per-token embeddings, assign each token a timestamp and duration, and group them into a fixed 30\,ms grid (duration-aware: each token contributes to all frames it spans). A per-frame MLP probe predicts binary mistake labels; predictions are merged into segments and evaluated with \texttt{mir\_eval.transcription} (segment $F1$, 2\,s tolerance). As this protocol relies on the model's tokenisation and onset/duration information, 
we therefore compare MAJEPPA only against the Aria baseline on UMP as other baselines use incomparable token schemes.

\textbf{Results}: As reported in Table~\ref{tab:evpmr}, the full MAJEPPA model attains the best score on Chopin, Technique, all regression tasks, and UMP EN/BM, while the $\mathcal{L}_{\text{piece}}{+}\mathcal{L}_{\text{cond}}$ variant is best on Pianist8 and UMP SI. MAJEPPA variants exceeds four baselines on Chopin, Technique, and the regression tasks, indicating that score-conditioned JEPA pre-training captures expressive and pedagogical structure not recovered by purely generative (Aria, Moonbeam) or masked-reconstruction (MusicBERT, CLaMP\,3) pre-training.
Aria's tokenizer has no explicit pedal tokens (sustain is folded into note durations), so the Aria baseline comparison does not isolate pedalling as a discrete cue.

The ablation reveals a division of labour among the three losses. $\mathcal{L}_{\text{piece}}$ alone is already competitive with the strongest baselines on most tasks. Adding $\mathcal{L}_{\text{cond}}$ benefits the tag-discriminative tasks (Pianist8, UMP SI), where samples with different but similar condition tags are brought closer according to $w_{ij}$. Adding $\mathcal{L}_{\text{self}}$ further reduces regression MAE and improves Chopin and UMP EN/BM, but slightly degrades Pianist8 and UMP SI: pulling within-performance segments together smooths over the segment-level cues these two tasks rely on.

\begin{figure}
    \centering
    \includegraphics[width=\linewidth]{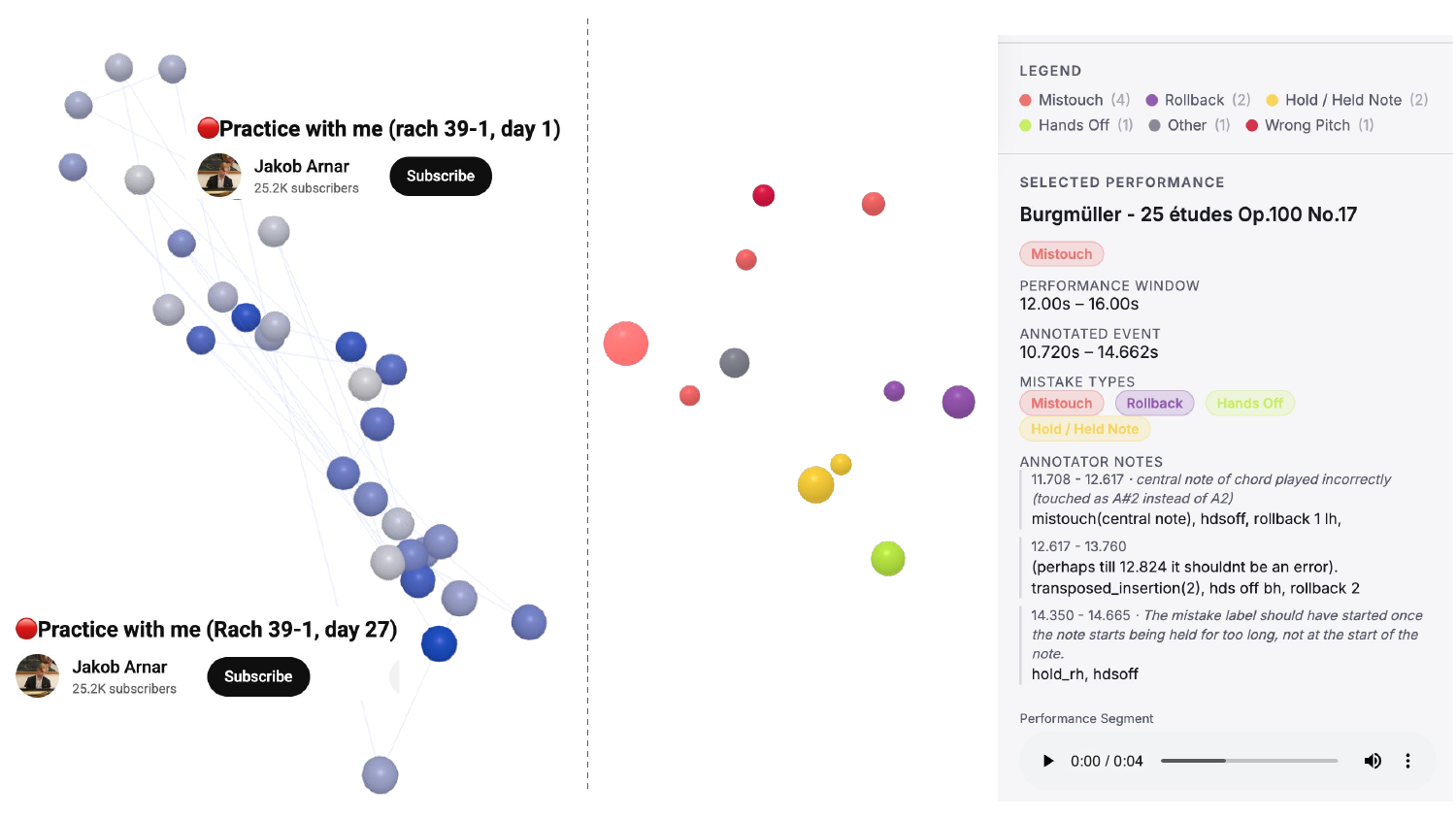}
    \caption{Samples from embedding viewer. \textbf{Left}: Practice trajectories for 30 days of Rachmaninov practice, coloured white (early) to blue (late). \textbf{Right}: UMAP of mistake regions for Burgmüller Étude Op.~100 No.~17 with annotations coloured by canonical mistake type, matching the interactive website.}
    \label{fig:embedding_viewer}
\end{figure}

\subsection{Generation and embedding space analysis}

Due to time constraints, we did not conduct a full-scale evaluation of the generation capabilities. However, we invite readers to visit our website\footnotemark[1] for sample outputs and interactive generation. In our demo website, we also showcase representative trajectories of PCA-reduced embeddings. 

\textbf{Practice progression.} On the session-averaged embeddings of 7 YouTube practice journeys, we measure Spearman rank correlations between session index (from earliest to latest session) and the trajectory's leading principal component. The mean correlation is $\rho = 0.41$ across PC1 and PC2, with trajectories exhibiting strong monotonic drift%
. Even in cases with moderate correlation, the latest-session points consistently concentrate at one end of each trajectory cloud (Figure~\ref{fig:embedding_viewer}), suggesting that MAJEPPA induces a directional notion of progress without explicit temporal supervision.

\textbf{Mistake topology.} Across the 25 Burgmüller études, we project mistake-annotated 4-second windows to 3D via UMAP, and further label 94 of the annotated regions with categorical mistake types (mistouch, wrong-pitch, rollback, …) drawn from independent human annotations in the SynMist project \cite{morsi2024simulating}. Colouring by mistake type (Figure 4, right) reveals contiguous within-piece subregions: timing-disruption mistakes (rollback / hands-off / hold) sit apart from pitch-deviation mistakes (mistouch / wrong-pitch), suggesting that MAJEPPA encodes the texture of a deviation rather than a single "is-a-mistake" axis.

\section{Conclusion}

We presented MAJEPPA, a dataset and framework for unified piano performance representations. We contribute ${\sim}$4,000 annotated recordings across six expertise levels and six recording contexts, spanning from child beginners to concert virtuosi. We use the dataset to train an LLM-JEPA model for symbolic music using a joint objective combining score-conditioned generation, cross-condition clustering, and within-performance alignment. 
MAJEPPA establishes a new representation space for music education, where performance quality, progression, and technique can be modelled, compared, and guided in a unified and data-driven manner.

\section{Ethics Statement}
MAJEPPA is built from publicly available YouTube solo-piano recordings for research purposes; because these recordings are publicly accessible, we did not obtain individual consent from uploaders. The corpus includes children's practice and showcase videos and commercial or concert recordings by professional pianists. However, public availability does not constitute a waiver of the uploaders' or performers' underlying rights. To mitigate this, we release only derived MIDI and coarse descriptive tags (expertise and recording context), not raw audio or personally identifying metadata beyond what is already public, and we will respond promptly to any request from an uploader, performer, or rights holder to remove specific content from the dataset. Additionally, we intend the resource for research on performance understanding and music education rather than for replacing teachers or performers. 

\section{Acknowledgments}
This work is supported by the UKRI Centre for Doctoral Training in Artificial Intelligence and Music. J. Zhou is a research student supported jointly by the China Scholarship Council and Queen Mary University of London.

\bibliography{ISMIRtemplate, ref}

@inproceedings{huang2025llmjepa,
        AUTHOR = {H. Huang and Y. LeCun and R. Balestriero},
        TITLE = {{LLM}-{JEPA}: Large Language Models Meet Joint Embedding Predictive Architectures},
        booktitle={Proceedings of the 14th International Conference on Learning Representations ({ICLR})},
        YEAR = {2026}
        }

@INPROCEEDINGS{bradshaw2025aria,
        AUTHOR = {L. Bradshaw and H. Fan and A. Spangher and S. Biderman and S. Colton},
        TITLE = {Scaling Self-Supervised Representation Learning for Symbolic Piano Performance},
        BOOKTITLE = {Proceedings of the 26th International Society for Music Information Retrieval Conference (ISMIR)},
        pages = {451--459},
        YEAR = {2025}
        }

@misc{bradshaw2024ariaamt,
        title={Aria-AMT: Efficient and robust automatic piano transcription},
        author={Louis Bradshaw and EleutherAI},
        year={2024},
        url={https://github.com/EleutherAI/aria-amt},
}

@INPROCEEDINGS{hu2022lora,
        AUTHOR = {E. J. Hu and Y. Shen and P. Wallis and Z. Allen-Zhu and Y. Li and S. Wang and L. Wang and W. Chen},
        TITLE = {{LoRA}: Low-Rank Adaptation of Large Language Models},
        BOOKTITLE = {International Conference on Learning Representations (ICLR)},
        YEAR = {2022}
        }

@ARTICLE{su2024roformer,
        AUTHOR = {J. Su and Y. Lu and S. Pan and A. Murtadha and B. Wen and Y. Liu},
        TITLE = {{RoFormer}: Enhanced Transformer with Rotary Position Embedding},
        JOURNAL = {Neurocomputing},
        VOLUME = {568},
        PAGES = {127063},
        YEAR = {2024},
        DOI = {10.1016/j.neucom.2023.127063}
        }

@INPROCEEDINGS{morsi2025rach3,
        AUTHOR = {A. Morsi and S. Chiruthapudi and S. D. Peter and M. Pilkov and L. Bishop and A. Maezawa and X. Serra and C. E. Cancino-Chac\'{o}n},
        TITLE = {Enabling Empirical Analysis of Piano Performance Rehearsal with the {Rach3} {MIDI} Dataset},
        BOOKTITLE = {Proceedings of the 26th International Society for Music Information Retrieval Conference (ISMIR)},
        pages = {484--491},
        YEAR = {2025}
        }

@inproceedings{kim2025pianovam,
  title={{PianoVAM}: A Multimodal Piano Performance Dataset},
  author={Kim, Yonghyun and Park, Junhyung and Bae, Joonhyung and Kim, Kirak and Kwon, Taegyun and Lerch, Alexander and Nam, Juhan},
  booktitle={Proceedings of the 26th International Society for Music Information Retrieval Conference (ISMIR)},
  pages = {528--535},
  year={2025}
}

@INPROCEEDINGS{mullerschon2025playability,
        AUTHOR = {M. M\"{u}llerschon and A. Klapuri and M. Rodriguez and C. Cardin},
        TITLE = {Playability Prediction in Digital Guitar Learning Using Interpretable Student and Song Representations},
        booktitle = {Proceedings of the 26th International Society for Music Information Retrieval Conference (ISMIR)},
        PAGES = {631--637},
        YEAR = {2025}
        }

@INPROCEEDINGS{libricky2025saxophone,
        AUTHOR = {\v{S}. Lib\v{r}ick\'{y} and J. Haji\v{c}},
        TITLE = {Modeling the Difficulty of Saxophone Music},
        booktitle = {Proceedings of the 26th International Society for Music Information Retrieval Conference (ISMIR)},
        PAGES = {747--754},
        YEAR = {2025}
        }

@INPROCEEDINGS{hasseinbey2025guitar,
        author = {Z. Hassein-Bey and Y. Abbou and A. D'Hooge and M. Giraud and G. Guillemain and A. Jeanneau},
        title = {What Song Now? {Personalized} Rhythm Guitar Learning in Western Popular Music},
        booktitle = {Proceedings of the 26th International Society for Music Information Retrieval Conference (ISMIR)},
        pages = {296--302},
        year = {2025}
        }

@INPROCEEDINGS{choi2025flutist,
        AUTHOR = {J. Choi and T. Kwon and J. Nam},
        TITLE = {Predicting Flutist Onset Timing in Duet Performance: A Multimodal Analysis of Gesture and Breath Cues},
       booktitle = {Proceedings of the 26th International Society for Music Information Retrieval Conference (ISMIR)},
       pages = {100--106},
        YEAR = {2025}
        }

@INPROCEEDINGS{zhang2025llaqo,
  author={Zhang, Huan and Cheung, Vincent K.M. and Nishioka, Hayato and Dixon, Simon and Furuya, Shinichi},
  booktitle={ICASSP 2025 - 2025 IEEE International Conference on Acoustics, Speech and Signal Processing (ICASSP)}, 
  title={{LLaQo}: Towards a Query-Based Coach in Expressive Music Performance Assessment}, 
  year={2025},
  pages={1--5},
  doi={10.1109/ICASSP49660.2025.10890522}}

@INPROCEEDINGS{morsi2024simulating,
        AUTHOR = {A. Morsi and H. Zhang and A. Maezawa and S. Dixon and X. Serra},
        TITLE = {Simulating Piano Performance Mistakes for Music Learning},
        BOOKTITLE = {Proceedings of the 21st  Sound and Music Computing Conference (SMC)},
        pages={179--186},
        YEAR = {2024}
        }

@inproceedings{Zhang2024FromJudges,
    author = {Zhang, Huan and Liang, Jinhua and Dixon, Simon},
    booktitle = {Proceedings of the 25th International Society for Music Information Retrieval Conference (ISMIR)},
    title = {From Audio Encoders to Piano Judges: Benchmarking Performance Understanding for Solo Piano},
    year = {2024},
    pages = {511--519}
}

@MISC{evpmr,
        AUTHOR = {Anonymous},
        TITLE = {{EVPMR}: Evaluation of Piano {MIDI} Representations},
        NOTE = {Software available at \url{https://github.com/anusfoil/eval-piano-midi-repr}},
        YEAR = {2026}
        }

@INPROCEEDINGS{foscarin2020asap,
        AUTHOR = {F. Foscarin and A. McLeod and P. Rigaux and F. Jacquemard and M. Sakai},
        TITLE = {{ASAP}: A Dataset of Aligned Scores and Performances for Piano Transcription},
        BOOKTITLE = {Proceedings of the 21st International Society for Music Information Retrieval Conference (ISMIR)},
        pages = {534--541},
        YEAR = {2020}
        }

@inproceedings{zhang_atepp_2022,
    address = {Bengaluru,India},
    title = {{ATEPP}: A Dataset of Automatically Transcribed Expressive Piano Performance},
    shorttitle = {ATEPP},
    language = {en},
    urldate = {2024-10-27},
    booktitle = {Proceedings of the 23rd International Society for Music Information Retrieval Conference (ISMIR)},
    author = {Zhang, Huan and Tang, Jingjing and Rafee, Syed RM and Dixon, Simon and Fazekas, George and Wiggins, Geraint A.},
    year = {2022},
    pages = {446--453}
}

@inproceedings{yan_scoring_2024,
    title = {Scoring Time Intervals Using Non-Hierarchical Transformer for Automatic Piano Transcription},
    doi = {10.5281/zenodo.14877493},
    booktitle = {Proceedings of the 25th International Society for Music Information Retrieval Conference (ISMIR)},
    author = {Yan, Yujia and Duan, Zhiyao},
    year = {2024},
    pages = {973--980}
}

@inproceedings{hu2025compose,
  title={Compose with Me: Collaborative Music Inpainter for Symbolic Music Infilling},
  author={Hu, Zhejing and Liu, Yan and Chen, Gong and Yu, Bruce X. B.},
  booktitle={Proceedings of the AAAI Conference on Artificial Intelligence},
  volume={39},
  number={2},
  pages={1327--1335},
  year={2025},
  doi={10.1609/aaai.v39i2.32122},
}

@INPROCEEDINGS{benetos2012score,
        AUTHOR = {E. Benetos and A. Klapuri and S. Dixon},
        TITLE = {{Score-informed} transcription for automatic piano tutoring},
        BOOKTITLE = {Proceedings of the 20th European Signal Processing Conference (EUSIPCO)},
        pages={2153-2157},
        YEAR = {2012}
        }

@inproceedings{morsi2023sounds,
  title={Sounds Out of Pl{\"a}ce? {Score-Independent} Detection of Conspicuous Mistakes in Piano Performances},
  author={Morsi, Alia and Tatsumi, Kana and Maezawa, Akira and Fujishima, Takuya and Serra, Xavier},
  booktitle = {Proceedings of the International Society for Music Information Retrieval Conference (ISMIR)},
  pages={352--358},
  year={2023}
}

@misc{EVAR,
    title = {{EVAR}: Evaluation package for Audio Representations},
    url = {https://github.com/nttcslab/eval-audio-repr},
    year = {2022}
}

@inproceedings{Zhang2024HowDataset,
author = {Zhang, Huan and Cheung, Vincent and Nishioka, Hayato and Dixon, Simon and Furuya, Shinichi},
booktitle = {International Society for Music Information Retrieval (ISMIR) Late Breaking Demo (LBD)},
title = {How does the teacher rate? {Observations} from the {NeuroPiano} dataset},
year = {2024}
}

@inproceedings{parmar2021pianoskillsassessment,
      title={Piano Skills Assessment}, 
      author={Paritosh Parmar and Jaiden Reddy and Brendan Morris},
      year={2021},
      booktitle={2021 IEEE 23rd International Workshop on Multimedia Signal Processing (MMSP)}, 
      pages={1--5}
}

@article{lecun2022path,
  title   = {A Path Towards Autonomous Machine Intelligence},
  author  = {LeCun, Yann},
  journal = {Open Review},
  volume={62},
  number={1},
  pages={1--62},
  year = {2022}
}

@inproceedings{huang2020remi,
  title={Pop Music Transformer: Beat-based Modeling and Generation of Expressive Pop Piano Compositions},
  author={Huang, Yu-Siang and Yang, Yi-Hsuan},
  booktitle={Proceedings of the 28th ACM International Conference on Multimedia},
  year={2020},
  pages={1180--1188},
  numpages = {9},
  series = {MM '20},
  doi={10.1145/3394171.3413671}
}

@inproceedings{hsiao2021cpword,
  title={Compound Word Transformer: Learning to Compose Full-Song Music over Dynamic Directed Hypergraphs},
  author={Hsiao, Wen-Yi and Liu, Jen-Yu and Yang, Yi-Hsuan},
  booktitle={Proceedings of the AAAI Conference on Artificial Intelligence},
  volume={35},
  number={1},
  pages={178--186},
  year={2021}
}

@inproceedings{lenz2024pertok,
  title={{PerTok}: Expressive Encoding and Modeling of Symbolic Musical Ideas and Variations},
  author={Lenz, Julian and Mani, Anirudh},
  booktitle={Proceedings of the 25th International Society for Music Information Retrieval Conference (ISMIR)},
  year={2024},
  pages={981--988},
  location={San Francisco, CA},
 
}

@inproceedings{zeng2021musicbert,
  title={{MusicBERT}: Symbolic Music Understanding with Large-Scale Pre-Training},
  author={Zeng, Mingliang and Tan, Xu and Wang, Rui and Ju, Zeqian and Qin, Tao and Liu, Tie-Yan},
  booktitle={Findings of the Association for Computational Linguistics: ACL-IJCNLP 2021},
  pages={791--800},
  year={2021}
}

@inproceedings{assran2023self,
  title={Self-Supervised Learning from Images with a Joint-Embedding Predictive Architecture},
  author={Assran, Mahmoud and Duval, Quentin and Misra, Ishan and Bojanowski, Piotr and Vincent, Pascal and Rabbat, Michael and LeCun, Yann and Ballas, Nicolas},
  booktitle={2023 IEEE/CVF Conference on Computer Vision and Pattern Recognition (CVPR)}, 
  year={2023},
  pages={15619-15629}
}

@article{assran2025vjepa2selfsupervisedvideo,
      title={{V-JEPA} 2: Self-Supervised Video Models Enable Understanding, Prediction and Planning}, 
      author={Mido Assran and Adrien Bardes and David Fan and Quentin Garrido and Russell Howes and Mojtaba Komeili and Matthew Muckley and Ammar Rizvi and Claire Roberts and Koustuv Sinha and Artem Zholus and Sergio Arnaud and Abha Gejji and Ada Martin and Francois Robert Hogan and Daniel Dugas and Piotr Bojanowski and Vasil Khalidov and Patrick Labatut and Francisco Massa and Marc Szafraniec and Kapil Krishnakumar and Yong Li and Xiaodong Ma and Sarath Chandar and Franziska Meier and Yann LeCun and Michael Rabbat and Nicolas Ballas},
      year={2025},
      eprint={2506.09985},
      archivePrefix={arXiv},
      primaryClass={cs.AI},
      Journal={arXiv preprint arXiv:2506.09985}
}

@article{guo2025moonbeammidifoundationmodel,
      title={Moonbeam: A MIDI Foundation Model Using Both Absolute and Relative Music Attributes}, 
      author={Zixun Guo and Simon Dixon},
      year={2025},
      eprint={2505.15559},
      archivePrefix={arXiv},
      primaryClass={cs.SD},
      journal={arXiv preprint arXiv:2505.15559}, 
}

@inproceedings{pilataki2025jepa_amt,
  title={Self-Supervised Representation Learning with a {JEPA} Framework for Multi-instrument Music Transcription},
  author={Pilataki, Mary and Mauch, Matthias and Dixon, Simon},
  booktitle={Proceedings of the IEEE Workshop on Applications of Signal Processing to Audio and Acoustics (WASPAA)},
  year={2025},
  pages={1--5},
  doi={10.1109/WASPAA66052.2025.11230951}
}

@article{you2025pianist,
  title={Pianist Transformer: Towards Expressive Piano Performance Rendering via Scalable Self-Supervised Pre-Training},
  author={You, Hong-Jie and Shao, Jie-Jing and Yang, Xiao-Wen and Jia, Lin-Han and Guo, Lan-Zhe and Li, Yu-Feng},
  journal={arXiv preprint arXiv:2512.02652},
  year={2025}
}

@inproceedings{tuncay2025audio,
  title = {{Audio-JEPA}: Joint-Embedding Predictive Architecture for Audio Representation Learning},
  author = {Tuncay, Ludovic and Labb{\'e}, Etienne and Benetos, Emmanouil and Pellegrini, Thomas},
  booktitle = {ICME 2025},
  address = {Nantes, France},
  year = {2025},
}

@inproceedings{StemJEPA,
    archivePrefix = {arXiv},
    arxivId = {2408.02514},
    author = {Riou, Alain and Lattner, Stefan and Hadjeres, Ga{\"{e}}tan and Anslow, Michael and Peeters, Geoffroy},
    booktitle = {Proceedings of the 25th International Society for Music Information Retrieval Conference (ISMIR)},
    eprint = {2408.02514},
    paeges = {625--633},
    title = {{Stem-JEPA}: A Joint-Embedding Predictive Architecture for Musical Stem Compatibility Estimation},
    year = {2024}
}

@inproceedings{hachana2025using,
    title={Using a Joint-Embedding Predictive Architecture for Symbolic Music Understanding},
    author={Rafik Hachana and Bader Rasheed},
    booktitle={AI for Music Workshop},
    year={2025}
}

@INPROCEEDINGS{min2025pianobart,
  author={Liang, Xiao and Zhao, Zijian and Zeng, Weichao and He, Yutong and He, Fupeng and Wang, Yiyi and Gao, Chengying},
  booktitle={2024 IEEE International Conference on Multimedia and Expo (ICME)}, 
  title={{PianoBART}: Symbolic Piano Music Generation and Understanding with Large-Scale Pre-Training}, 
  year={2024},
  volume={},
  number={},
  pages={1-6},
  doi={10.1109/ICME57554.2024.10688332}}

@inproceedings{yuan2024chatmusicianunderstandinggeneratingmusic,
      title={{ChatMusician}: Understanding and Generating Music Intrinsically with {LLM}}, 
      author={Ruibin Yuan and Hanfeng Lin and Yi Wang and Zeyue Tian and Shangda Wu and Tianhao Shen and Ge Zhang and Yuhang Wu and Cong Liu and Ziya Zhou and Ziyang Ma and Liumeng Xue and Ziyu Wang and Qin Liu and Tianyu Zheng and Yizhi Li and Yinghao Ma and Yiming Liang and Xiaowei Chi and Ruibo Liu and Zili Wang and Pengfei Li and Jingcheng Wu and Chenghua Lin and Qifeng Liu and Tao Jiang and Wenhao Huang and Wenhu Chen and Emmanouil Benetos and Jie Fu and Gus Xia and Roger Dannenberg and Wei Xue and Shiyin Kang and Yike Guo},
      booktitle = {Findings of the Association for Computational Linguistics: ACL 2024},
      pages = {6252--6271},
      year={2024},
}

@article{lu2023musecocogeneratingsymbolicmusic,
      title={{MuseCoco}: Generating Symbolic Music from Text}, 
      author={Peiling Lu and Xin Xu and Chenfei Kang and Botao Yu and Chengyi Xing and Xu Tan and Jiang Bian},
      year={2023},
      eprint={2306.00110},
      archivePrefix={arXiv},
      primaryClass={cs.SD},
      journal={arXiv preprint arXiv:2306.00110}, 
}

@article{midibertpiano,
  title={{MidiBERT-Piano}: Large-scale Pre-training for Symbolic Music Understanding},
  author={Yi-Hui Chou and I-Chun Chen and Chin-Jui Chang and Joann Ching and Yi-Hsuan Yang},
  journal={Journal of Creative Music Systems ({JCMS})},
  volume={8},
  number={1},
  doi={https://doi.org/10.5920/jcms.1064},
  year={2024}
}

@INPROCEEDINGS{wu2025clamp3universalmusic,
    AUTHOR = {Shangda Wu and Zhancheng Guo and Ruibin Yuan and Junyan Jiang and SeungHeon Doh and Gus Xia and Juhan Nam and Xiaobing Li and Feng Yu and Maosong Sun},
    TITLE = {{CL}a{MP} 3: Universal Music Information Retrieval Across Unaligned Modalities and Unseen Languages},
    BOOKTITLE = {Findings of the Association for Computational Linguistics: ACL 2025},
    PAGES = {2605--2625},
    YEAR = {2025}
}

@article{ramoneda2023cipi,
    title={Combining piano performance dimensions for score difficulty classification},
    author={Ramoneda, Pedro and Jeong, Dasaem and Eremenko, Vsevolod and Tamer, Nazif Can and Miron, Marius and
  Serra, Xavier},
    journal={Expert Systems with Applications},
    ISSN={0957-4174},
    Volumn={238},
    Pages={121776},
    year={2024}
  }

@inproceedings{tang2023pianistid,
    title={Pianist Identification Using Convolutional Neural Networks},
    author={Tang, Jingjing and Wiggins, Geraint and Fazekas, Gyorgy},
    booktitle={Proceedings of the 4th International Symposium on the Internet of Sounds (IS2)},
    year={2023}
  }

@inproceedings{chou2026laddersym,
    title={LadderSym: A Multimodal Interleaved Transformer for Music Practice Error Detection},
    author={Chou, Benjamin Shiue-Hal and Jajal, Purvish and Eliopoulos, Nick John and Davis, James C. and  Thiruvathukal, George K. and Yun, Kristen Yeon-Ji and Lu, Yung-Hsiang},
    booktitle={The Fourteenth International Conference on Learning Representations (ICLR)},
    year={2026}
  }

@inproceedings{chang2025RUMAA,
  title = {{RUMAA}: Repeat-Aware Unified Music Audio Analysis for Score-Performance Alignment, Transcription, and Mistake Detection},
  shorttitle = {RUMAA},
  booktitle = {2025 IEEE Workshop on Applications of Signal Processing to Audio and Acoustics (WASPAA)},
  author = {Chang, Sungkyun and Dixon, Simon and Benetos, Emmanouil},
  year={2025},
  date = {2025-10},
  pages = {1--5},
  issn = {1947-1629},
  doi = {10.1109/WASPAA66052.2025.11230990},
  eventtitle = {2025 {{IEEE Workshop}} on {{Applications}} of {{Signal Processing}} to {{Audio}} and {{Acoustics}} ({{WASPAA}})}
}

@article{article,
author = {Gelbukh, Alexander and {\'{A}}lvarez, Daniel and Kolesnikova, Olga and Chanona-Hern{\'{a}}ndez, Liliana and Sidorov, Grigori},
doi = {10.13053/cys-28-1-4903},
journal = {Computacion y Sistemas},
pages = {85--98},
title = {{Multi-Instrument Based N-Grams for Composer Classification Task}},
volume = {28},
year = {2024}
}

@article{Wang2021Audio-basedMechanism,
author = {Wang, Weiqing and Pan, Jin and Yi, Hua and Song, Zhanmei and Li, Ming},
doi = {10.1109/TASLP.2021.3061267},
issn = {23299304},
journal = {IEEE/ACM Transactions on Audio Speech and Language Processing},
pages = {1119--1133},
title = {Audio-Based Piano Performance Evaluation for Beginners with Convolutional Neural Network and Attention Mechanism},
volume = {29},
year = {2021}
}

@inproceedings{Jiang2023ExpertFeedback,
author = {Jiang, Yucong},
booktitle = {Proceedings of the 24th International Society for Music Information Retrieval (ISMIR)},
title = {Expert and Novice Evaluations of Piano Performances: Criteria for Computer-Aided Feedback},
pages = {367--374},
year = {2023}
}

%
%
%
%

\end{document}